\documentclass[10pt]{iopart}
\expandafter\let\csname equation*\endcsname\relax
\expandafter\let\csname endequation*\endcsname\relax

\usepackage{hyperref}
\hypersetup{
  colorlinks,
  citecolor={cyan!50!black},
  linkcolor={orange!60!black},
  urlcolor=violet,
}
\usepackage{orcidlink}
\usepackage{graphicx}
\usepackage{amsmath,amssymb}
\usepackage{physics}
\usepackage{mathtools}  %
\usepackage{tensor}
\usepackage[caption=false]{subfig}
\usepackage{bm} %
\usepackage{commath}  %
\usepackage[binary-units]{siunitx}
\usepackage[capitalise]{cleveref}
\usepackage[normalem]{ulem} %
\usepackage{csquotes}

\crefname{section}{Sec.}{Sec.}
\Crefname{section}{Section}{Sections}

\newcommand{\ccite}[1]{Ref.~\cite{#1}}
\newcommand{\ccites}[1]{Refs.~\cite{#1}}
\newcommand{\Ccite}[1]{Reference~\cite{#1}}

\newcommand{\tikzinput}[1]{
  \includegraphics{#1.pdf}
}

\newcommand{\defeq}{\coloneqq}

\newcommand{\dgF}{\mathcal{F}}
\newcommand{\dgFi}[1]{\tensor{\dgF}{_{\!{#1}}^i}}

\newcommand{\bas}{\psi}
\newcommand{\lagr}{\ell}
\newcommand{\jac}{\mathrm{J}}
\newcommand{\invjac}{{(\jac^{-1})}}
\newcommand{\surf}[1]{{#1}^\Sigma}

\newcommand{\dgA}{\mathcal{A}}
\newcommand{\dgM}{M}
\newcommand{\dgD}{D}
\newcommand{\dgMD}{M\!\!D}
\newcommand{\dgL}{L}
\newcommand{\dgML}{M\!\!L}

\newcommand{\opA}{\dgA}

\newcommand{\grd}[1]{\underline{#1}}

\newcommand{\penaltyparam}{C}

\newcommand{\displ}{u}
\newcommand{\strain}{S}
\newcommand{\stress}{T}
\newcommand{\epot}{U}
\newcommand{\constrel}{Y}

\newcommand{\lossang}{\phi}
\newcommand{\sub}{\mathrm{sub}}
\newcommand{\coat}{\mathrm{coat}}
\newcommand{\poiss}{\sigma}
\newcommand{\youngs}{Y}

\newcommand{\kb}{k_\mathrm{B}}

\newcommand{\algaas}{AlGaAs}

\newcommand{\spectre}{\texttt{SpECTRE}}
\newcommand{\dealii}{\texttt{deal.ii}}

\begin{document}

\title{High-accuracy numerical models of Brownian thermal noise in thin mirror coatings}

\author{%
Nils L Vu$^{1,4}$\footnote{Corresponding author}\,\orcidlink{0000-0002-5767-3949}, %
Samuel Rodriguez${^{2,3}}$\,\orcidlink{0000-0002-1879-8810}, %
Tom W\l{}odarczyk$^1$\,\orcidlink{0000-0003-0005-348X}, %
Geoffrey Lovelace$^2$\,\orcidlink{0000-0002-7084-1070}, %
Harald P Pfeiffer$^1$\,\orcidlink{0000-0001-9288-519X}, %
Gabriel S Bonilla$^2$\,\orcidlink{0000-0003-4502-528X}, %
Nils Deppe$^4$\,\orcidlink{0000-0003-4557-4115}, %
Fran\c{c}ois H\'{e}bert$^4$\,\orcidlink{0000-0001-9009-6955}, %
Lawrence E Kidder$^5$\,\orcidlink{0000-0001-5392-7342}, %
Jordan Moxon$^4$\,\orcidlink{0000-0001-9891-8677}, and %
William Throwe$^5$\,\orcidlink{0000-0001-5059-4378}%
}

\vspace{1pc}
\address{$^1$ Max Planck Institute for Gravitational Physics
(Albert Einstein Institute), Am Mühlenberg 1, Potsdam 14476, Germany}
\address{$^2$ Nicholas and Lee Begovich Center for
Gravitational-Wave Physics and Astronomy, California State University
Fullerton, Fullerton, California 92831, USA}
\address{$^3$ Department of Physics and Astronomy,
University of Mississippi, University, Mississippi 38677, USA}
\address{$^4$ Theoretical Astrophysics, Walter Burke
Institute for Theoretical Physics, California Institute of Technology, Pasadena,
California 91125, USA}
\address{$^5$ Cornell Center for Astrophysics and Planetary
Science, Cornell University, Ithaca, New York 14853, USA}

\ead{owls@nilsvu.de}

\date{\today}

\begin{abstract}
Brownian coating thermal noise in detector test masses is limiting the
sensitivity of current gravitational-wave detectors on Earth.
Therefore, accurate numerical models can inform the ongoing effort to minimize
Brownian coating thermal noise in current and future gravitational-wave
detectors.
Such numerical models typically require significant computational resources and
time, and often involve closed-source commercial codes. In contrast, open-source
codes give complete visibility and control of the simulated physics, enable
direct assessment of the numerical accuracy, and support the reproducibility of
results.
In this article, we use the open-source \spectre{} numerical relativity code and
adopt a novel discontinuous Galerkin numerical method to model Brownian coating
thermal noise.
We demonstrate that \spectre{} achieves significantly higher accuracy than a
previous approach at a fraction of the computational cost.  Furthermore, we
numerically model Brownian coating thermal noise in multiple sub-wavelength
crystalline coating layers for the first time.
Our new numerical method has the potential to enable fast exploration of
realistic mirror configurations, and hence to guide the search for optimal
mirror geometries, beam shapes and coating materials for gravitational-wave
detectors.
\end{abstract}

\vspace{2pc}
\noindent{\it Keywords}:
gravitational-wave detectors,
Brownian coating thermal noise,
numerical simulation,
discontinuous Galerkin methods

\submitto{\CQG}

\setcounter{footnote}{0}
\makeatletter
\long\def\@makefntext#1{\parindent 1em\noindent 
 \makebox[1em][l]{\footnotesize\rm$\m@th{^\arabic{footnote}}$}%
 \footnotesize\rm #1}
\def\@makefnmark{\hbox{${^\arabic{footnote}}\m@th$}}
\def\@thefnmark{\arabic{footnote}}
\makeatother

\newlength{\figwidth}
\setlength{\figwidth}{.5\columnwidth}

\section{Introduction}

Brownian coating thermal noise is the limiting noise source for
current-generation, ground-based gravitational-wave detectors in their most
sensitive frequency bands. For instance, following the A+ upgrade anticipated
for completion in the mid 2020s, the Laser Intererometer Gravitational-Wave
Observatory (LIGO) detector noise is dominated by Brownian coating thermal noise
at frequencies $f\sim \SI{100}{\hertz}$~\cite{aplusNoise}. This noise arises from
thermal fluctuations in the reflective coatings of the detectors' test masses~\cite{Cole2013-wp}.

Therefore, a reduction of the Brownian coating thermal noise directly increases a
detector's
sensitivity and thus its astronomical reach. Theoretical models of Brownian
coating thermal noise are important for working toward this goal. Thermal noise
modeling typically follow the approach pioneered by
Levin~\cite{Levin:1997kv}, which computes the thermal noise in terms of an
auxiliary elasticity calculation using the fluctuation-dissipation theorem~\cite{callen1951irreversibility,bernard1959fdt,kubo1966fluctuation}.
While an approximate analytic solution is well known in the limit where coating
thickness and edge effects can be neglected, numerical calculations of thermal
noise are necessary to study effects that arise from the finite test-mass size,
the finite coating thickness, and from crystalline materials.

In this article we calculate Brownian coating thermal noise
by numerically solving the auxiliary linear elasticity problem.
Such numerical simulations typically adopt a conventional finite-element
approach, as some of the authors did in \ccite{Lovelace2017xyf}. These methods
are widely used, but achieving high accuracy with them can require significant
computational resources and time, because of their relatively slow rates of
convergence.

For the first time to our knowledge, we apply a discontinuous Galerkin (DG)
method to model Brownian coating thermal noise.
DG methods are well suited to this problem because they can retain high-order
convergence in the presence of discontinuities, which arise at the interfaces
between the mirror substrate and its reflective coatings.
In this article, we extend the DG method for elliptic equations presented in
\ccite{dgscheme} to problems with discontinuous material properties. With
this extension, our method converges exponentially with resolution, allowing
us to solve coating thermal noise problems numerically at high accuracy using
considerably less computational resources and time than conventional
finite-element methods.

We implement the numerical method and the elastostatic equations in
\spectre{}~\cite{spectre}, a new open-source numerical relativity code.
While \spectre{}'s primary aim is to
model merging black holes and neutron stars, the elliptic solver needed to
construct initial data for such simulations is also very well positioned to
solve the DG-discretized elastostatics equations for thermal noise modeling~\cite{ellsolver, dgscheme}. As an open-source
code, our approach has advantages compared to the closed-source and commercial solutions that
are often adopted: we can directly control the physics incorporated in the
calculation, we can assess the accuracy and convergence rate of our simulations
in a straightforward way, and our results are reproducible with publicly
available software.
Our code also benefits from \spectre{}'s task-based
parallelism approach, implemented using the \texttt{Charm++}~\cite{charmpp}
library, enabling our code
to efficiently scale to large numbers of compute cores~\cite{Kidder2017-nz}.

This article is organized as follows. \Cref{sec:methods}
summarizes the elastic problem to be solved and presents the
discontinuous Galerkin numerical method. \Cref{sec:results} presents our
results using this method to model thermal noise in cylindrical mirrors
with thin coatings.
We discuss our results and future work in \cref{sec:conclusion}.

\section{Methods}\label{sec:methods}

In this section, we formulate the auxiliary elasticity problem based on
\ccites{Levin:1997kv,Lovelace2017xyf}, discretize it with the discontinuous
Galerkin scheme developed in \ccite{dgscheme}, and outline the numerical method
we employ to solve the discretized problem with the \spectre{}
code~\cite{ellsolver}.
\Cref{sec:numflux} details a novel extension of this method to handle
discontinuous material properties at layer interfaces.

\subsection{Auxiliary elasticity problem}

We consider a gravitational-wave detector
that measures the position of a test mass with a laser beam with a Gaussian
intensity profile
\begin{equation}\label{eq:laserbeam}
  p(r) = \frac{1}{\pi r_0^2} e^{-r^2/r_0^2}
  \text{.}
\end{equation}
Here, $r$ is the cylindrical radial coordinate from the center of the beam
with width~$r_0$.
The intensity profile is normalized so that
\begin{equation}
  \int_0^{2\pi} \! \dd{\phi} \int_0^\infty \! \dd{r} \, r p(r) = 1
  \text{.}
\end{equation}
The laser beam effectively measures a weighted average~$q$ of the
displacement~$Z$ of the test mass surface,
\begin{equation}
  q(t) = \int_0^{2\pi} \! \dd{\phi} \int_0^R \! \dd{r} \, r p(r,\phi) Z(r,\phi,t)
  \text{.}
\end{equation}

As shown by Levin~\cite{Levin:1997kv}, Brownian thermal noise can be
calculated from the energy dissipated in an auxiliary elastic problem.
Specifically, to compute the thermal noise at frequency~$f$, one applies an
oscillating pressure to the face of the mirror with frequency~$f$, with a
pressure distribution profile~$p(r)$ equal to the beam intensity, and with an amplitude~$F_0$.  In this auxiliary problem,
the energy~$W_\mathrm{diss}$ will be dissipated in each cycle of the oscillation. The
fluctuation-dissipation theorem relates this dissipated energy~$W_\mathrm{diss}$ to
the thermal noise, specifically to the power spectral density $S_{q}$ associated
with $q$,\footnote{See, e.g., Eq.~(11.90) in \ccite{BlandfordThorne}.}
\begin{equation}\label{eq:thermalnoise}
  S_q = \frac{2 \kb{} T}{\pi^2 f^2} \frac{W_\mathrm{diss}}{F_0^2}
  \text{,}
\end{equation}
where $T$ is the mirror temperature and $\kb{}$ is Boltzmann's constant. Because
$W_\mathrm{diss} \propto F_0^2$, it follows that $S_q$ does not depend on the overall amplitude~$F_0$.

For frequencies $f\sim \SI{100}{\hertz}$ much lower than the resonant frequencies
$f \sim 10^4\si{\hertz}$ of the test-mass materials, the dissipated power can be
computed using the quasistatic approximation. In this approximation, a static
pressure is applied to the mirror with amplitude~$F_0$ and profile~$p(r)$, and
the dissipated energy can be written as
\begin{equation}\label{eq:wdiss}
  W_\mathrm{diss} = \epot \lossang,
\end{equation}
where $\epot$ is the potential energy stored in the deformation of
the test-mass and $\lossang$ is the material's loss angle
determined by the material's imaginary, dissipative elastic moduli.

Therefore, our goal in this article is to solve the equations of elastostatics
for the deformation of the test mass,
\begin{equation}\label{eq:Newton}
  \nabla_i \stress^{ij} = f^j(\bm{x}),
\end{equation}
when its surface is subjected to an applied pressure with profile~$p(r)$.
Here, $\stress^{ij}$ is the \emph{stress} and we adopt the Einstein summation convention so
that repeated tensor indices are summed over. The source~$f^j$ is the force
density acting on each volume element of the mirror as a function of position
$\bm{x}$, which vanishes in our situation, $f^j=0$.  The pressure acting on the
external surface of the test mass will be represented by suitable boundary
conditions.

\Cref{eq:Newton}
is an equation for the \emph{displacement vector field}~$\displ^i(\bm{x})$, which
describes the deformation of the elastic material as a function of the undeformed coordinates. The symmetric part of the
gradient of the displacement vector field is the \emph{strain}
\begin{equation}\label{eq:strain}
  \strain_{kl} = \nabla_{(k} \displ_{l)}
  \text{.}
\end{equation} For sufficiently small $F_0$, the strain is proportional to the
applied stress,
\begin{equation}\label{eq:Hooke}
  \stress^{ij} = -\constrel^{ijkl} \strain_{kl}
  \text{,}
\end{equation}
where the \emph{constitutive relation}~$\constrel^{ijkl}(\bm{x})$ captures the elastic
properties of the material in the linear regime.  The constitutive relation
is symmetric on its first two indices, on its last two indices, and under exchange of the first pair of indices with the second pair of indices.

Inserting \cref{eq:Hooke,eq:strain} into \cref{eq:Newton} yields the
equations of linear elasticity,
\begin{equation}\label{eq:elasticity_eqns}
  -\nabla_i \constrel^{ijkl} \nabla_{(k} \displ_{l)} = f^j(\bm{x})
  \text{,}
\end{equation}
which we will solve numerically.

We consider materials with either amorphous or cubic-crystalline constitutive
relations. The coating may consist of only a single material, or of multiple
layers with different materials. The amorphous constitutive relation is
isotropic and homogeneous,
\begin{equation}\label{eq:constrel_iso}
  \constrel^{ijkl} = \lambda \delta^{ij}\delta^{kl} + \mu
  \left(\delta^{ik}\delta^{jl} + \delta^{il}\delta^{jk}\right)
  \text{,}
\end{equation}
with \emph{Lamé parameter}~$\lambda$ and \emph{shear modulus}~$\mu$.\footnote{The
Lamé parameter can also be replaced by the \emph{bulk modulus}~$K=\lambda +
2\mu/3$. Alternatively, the two parameters can be replaced by the
\emph{Young's modulus}~$\youngs=9K\mu/(3K+\mu)=\mu(3\lambda+2\mu)/(\lambda+\mu)$ and the
\emph{Poisson ratio}~$\poiss=(3K-2\mu)/(2(3K+\mu))=\lambda/(2(\lambda+\mu))$.}
A cubic-crystalline material is characterized by the constitutive relation
\begin{equation}\label{eq:constrel_cryst}
  \constrel^{ijkl} =
  \begin{cases}
  c_{11} & \mathrm{for}\; i=j=k=l \\
  c_{12} & \mathrm{for}\; i=j,k=l,i \neq k \\
  c_{44} & \mathrm{for}\; i=k,j=l,i \neq j \;\mathrm{or}\; i=l,j=k,i\neq j
\end{cases}
\end{equation}
where $c_{11}$, $c_{12}$ and $c_{44}$ are three independent material parameters.
The constitutive relation in \cref{eq:elasticity_eqns} is composed by
discontinuously choosing either \cref{eq:constrel_iso} or
\cref{eq:constrel_cryst} in each layer of the material.

After solving the linear elasticity equations, \cref{eq:elasticity_eqns}, the
potential energy is evaluated by an integral over the volume of the material,
\begin{equation}
  \epot = -\frac{1}{2}\int_V dV \strain_{ij} \stress^{ij}
  \text{.}
\end{equation}

For a material with a thin, reflective coating
with different elastic properties than the substrate, the dissipated
energy, \cref{eq:wdiss}, decomposes as~\cite{Harry2002-tv}
\begin{equation}
  W_\mathrm{diss} = \epot_\sub \, \lossang_\sub + \epot_\coat \, \lossang_\coat,
\end{equation}
where $\epot_\sub$ and $\lossang_\sub$ are the potential energy and loss angle
of the substrate, respectively, while $\epot_\coat$ and $\lossang_\coat$ are the
potential energy and the loss angle of the coating. Note that a material can
also have different loss angles associated with the different independent
elastic moduli of a material~\cite{Hong2013}. We do not consider further decompositions of the
elastic potential energy in this article, but note that such quantities can
straightforwardly be extracted from our simulations. The different loss angles
only affect the computation of the thermal noise by \cref{eq:thermalnoise} from
the elastic potential energy extracted from our simulations, but they do not
affect our simulations or our numerical method in any way.

An approximate analytic solution exists for amorphous materials in the limit where the coating
thickness~$d$ is small compared to both the size of the mirror and the width~$r_0$ of the
pressure profile. The approximate coating thermal
noise is \footnote{See Eq.~(22) in \ccite{Harry2002-tv}, where $w=\sqrt{2} \,
r_0$, $\lossang_\parallel=\lossang_\lossang=\lossang_\coat$, and we consider
only the coating contribution.}
\begin{align}\label{eq:approx_analytic}
  S_q^{\,\coat} = &\frac{\kb{} T}{\pi^2 f}\frac{1-\poiss_\sub^2}{r_0\youngs_\sub}\frac{d}{r_0}
  \frac{\lossang_\coat}{\youngs_\sub \youngs_\coat (1-\poiss_\coat^2) (1-\poiss_\sub^2)} \nonumber \\
  &\times \bigl( \begin{aligned}[t]
    &\youngs_\coat^2 (1+\poiss_\sub)^2(1-2\poiss_\sub)^2 + \\
    &\youngs_\sub^2 (1+\poiss_\coat)^2 (1-2\poiss_\coat) \bigr)
    \text{.}
  \end{aligned}
\end{align}

\subsection{Discontinuous Galerkin discretization}\label{sec:dgscheme}

We employ the discontinuous Galerkin (DG) scheme detailed in \ccite{dgscheme} to
discretize the elasticity problem, \cref{eq:elasticity_eqns}. We summarize the
discretization scheme in this section, and extend it to problems with
discontinuous material properties.

\subsubsection{Domain decomposition}

\begin{figure}
  \centering
  \subfloat[Domain\protect\label{fig:domain}]{
    \tikzinput{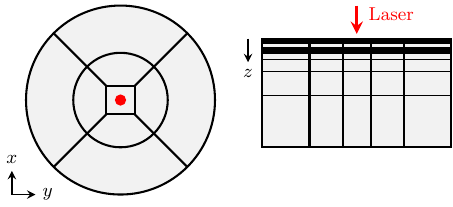}
  }\\
  \subfloat[Element\protect\label{fig:element}]{
    \tikzinput{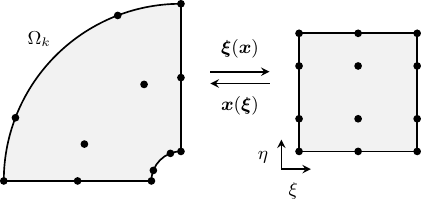}
  }
  \caption{
    \label{fig:domain_and_elements}
    \emph{Top:} Geometry of our layered cylindrical domain, with the laser beam
    indicated in red. Four wedge-shaped elements envelop a cuboid. Another set
    of wedges extends to the outer radius of the cylinder. In $z$~direction the
    cylinder is partitioned into layers that can have different material
    properties (black and gray). The substrate layer has a logarithmic
    coordinate map in $z$~direction and is split in two twice in this example
    (thin horizontal lines). \emph{Bottom:} The coordinate
    transformation~$\bm{\xi}(\bm{x})$ maps an element to a reference
    cube~$\bm{\xi}\in[-1,1]^3$ with logical coordinate-axes~$\bm{\xi}=(\xi,\eta,\zeta)$.
    In this example we chose $N_{k,\xi}=3$ and $N_{k,\eta}=4$ LGL collocation
    points along $\xi$ and $\eta$, respectively.}
\end{figure}

We simulate a cylindrical mirror in three dimensions with radius $R$ and height
$H$. The cylinder axis coincides with the $z$~axis of our coordinates, and the
plane $z=0$ represents the surface of the mirror on which the external pressure~$p(r)$ is applied.
We decompose the cylindrical domain $\Omega = [0,R] \times
[0,2\pi) \times [0,H]$ into a set of nonoverlapping elements~$\Omega_k \subset
\Omega$ shaped like deformed cubes, as illustrated in \cref{fig:domain} ($h$~refinement). Each
element carries a coordinate map from the Cartesian coordinates $\bm{x} \in
\Omega_k$, in which the elasticity equations~\eqref{eq:elasticity_eqns} are
formulated, to \emph{logical} coordinates $\bm{\xi} \in [-1, 1]^3$
representing the reference cube, as illustrated in \cref{fig:element}. The
coordinate map to the reference cube is characterized by its Jacobian,
\begin{equation}
  \jac^i_j = \pdv{{x^i}}{{\xi^j}}
\end{equation}
with determinant~$\jac$ and inverse~$\invjac^j_i = \partial \xi^j / \partial
x^i$. On the reference cube we choose a set of $N_{k, i}$ Legendre-Gauss-Lobatto
(LGL) collocation points in each dimension~$i$ ($p$~refinement).

Fields are represented numerically by their values at the collocation points. We
denote the set of discrete values for the displacement vector field $\displ^i$ within
an element~$\Omega_k$ as
\begin{equation}
  \grd{\displ}^{i,(k)}=(\displ^{i,(k)}_1,\ldots,\displ^{i,(k)}_{N_k})
  \text{,}
\end{equation}
and the collection of discrete displacement vector field values over \emph{all}
elements as~$\grd{\displ}^i$. The values at the collocation points within an element
define a three-dimensional Lagrange interpolation,
\begin{equation}\label{eq:field_expansion}
  \displ^{i,(k)}(\bm{x}) \defeq \sum_{p=1}^{N_k} \displ^i_p \bas_p(\bm{\xi}(\bm{x}))
  \quad \text{with} \quad \bm{x} \in \Omega_k
  \text{,}
\end{equation}
where the basis functions~$\bas_p(\bm{\xi})$ are products of Lagrange
polynomials,
\begin{equation}\label{eq:lagrprod}
  \bas_p(\bm{\xi}) \defeq \prod_{i=1}^3 \lagr_{p_i}(\xi^i)
  \quad \text{with} \quad \bm{\xi} \in [-1,1]^3
  \text{,}
\end{equation}
based on the collocation points in the three logical directions of the element~$\Omega_k$. Since
\cref{eq:field_expansion,eq:lagrprod} are local to each element, fields
over the entire domain are discontinuous across element boundaries.

\subsubsection{DG residuals}

To formulate the elasticity equations in first-order form for the DG
discretization, we use the symmetric strain~$\strain_{kl}$ as auxiliary variable.
Following \ccite{dgscheme}, we first compute the discrete auxiliary
variables on the computational grid as
\begin{equation}
  \grd{\strain}_{kl} = \dgD_{(k} \cdot \grd{\displ}_{l)} + \dgL\cdot((n_{(k}\grd{\displ}_{l)})^* - n_{(k} \grd{\displ}_{l)})
  \text{,}
\end{equation}
where we make use of the discrete differentiation matrix $\dgD_i \defeq
\dgM^{-1}\dgMD_i$, the mass matrix
\begin{align}
  \label{eq:massmat}
  \dgM_{pq} &= \int_{[-1,1]^3} \bas_p(\bm{\xi}) \bas_q(\bm{\xi}) \, \jac \dd{^3\xi}
  \text{,}
\intertext{the stiffness matrix}
  \label{eq:stiffmat}
  \dgMD_{i,pq} &= \int_{[-1,1]^3} \bas_p(\bm{\xi}) \, \pdv{\bas_q}{{\xi^j}}\!(\bm{\xi})
  \, \invjac^j_i \, \jac \dd{^3\xi}
  \text{,}
\intertext{the lifting operator}
  \label{eq:liftop}
  \dgML_{pq} &= \int_{[-1,1]^2} \bas_p(\bm{\xi})\bas_q(\bm{\xi})
  \, \surf{\jac} \dd{^2\xi}
  \text{,}
\end{align}
and $\dgL \defeq \dgM^{-1}\dgML$ on the element~$\Omega_k$~\cite{dgscheme}.
The integral in \cref{eq:liftop} is over the boundary of the element,
$\partial\Omega_k$, where $n_i$ is the outward-pointing unit normal one-form and
$\surf{\jac}$ is the surface Jacobian.
The symbol $\cdot$ emphasizes matrix multiplication with the field values over
the computational grid of the element.
In a second step, we compute the DG residuals in strong form~\cite{dgscheme},
\begin{align}\label{eq:dg_residuals}
  -\dgMD_i \cdot \constrel^{ijkl} \grd{\strain}_{kl} - \dgML \cdot ((n_i \constrel^{ijkl} \grd{\strain}_{kl})^*
  - n_i \constrel^{ijkl} \grd{\strain}_{kl})
  = \dgM \cdot \grd{f}^j
  \text{,}
\end{align}
which represent the set of algebraic equations for the values~$\grd{\displ}^i$ of the
displacement vector field on the computational grid that we solve numerically.

\subsubsection{Numerical flux}\label{sec:numflux}

The quantities $(n_{(k}\grd{\displ}_{l)})^*$ and $(n_i \constrel^{ijkl} \grd{\strain}_{kl})^*$ in
\cref{eq:dg_residuals} denote a numerical flux that couples grid points across
nearest-neighbor element boundaries. We employ the generalized internal-penalty
numerical flux developed in \ccite{dgscheme}, with one notable extension.
Contrary to \ccite{dgscheme} we allow neighboring elements to define different
constitutive relations, meaning $\constrel^{ijkl}(\bm{x})$ can be double-valued
on shared element boundaries.\footnote{In the language of \ccite{dgscheme} we
allow the fluxes $\dgFi{\alpha}[u_A,v_A;\bm{x}]$ to be double-valued on shared
element boundaries.} Therefore, we define the quantity
\begin{equation}\label{eq:constrel_avg}
  \constrel^{ijkl}_* = \frac{1}{2}\left(\constrel^{ijkl}_\mathrm{int} + \constrel^{ijkl}_\mathrm{ext}\right)
  \text{,}
\end{equation}
where \enquote{int} denotes the \emph{interior} side of an element's shared boundary
with a neighbor, and \enquote{ext} denotes the \emph{exterior} side, i.e.\ the
neighbor's side. With this quantity we can define the numerical flux
\begin{subequations}\label{eq:numflux}
\begin{align}
  (n_{(k} \grd{\displ}_{l)})^* = \frac{1}{2}\Bigl[
    &n_{(k}^\mathrm{int} \grd{\displ}_{l)}^\mathrm{int} -
    n_{(k}^\mathrm{ext} \grd{\displ}_{l)}^\mathrm{ext}\Bigr]
    \label{eq:numfluxaux}\text{,}\\
  (n_i \constrel^{ijkl} \grd{\strain}_{kl})^* =
    \frac{1}{2}\Bigl[
    &n_i^\mathrm{int} \constrel^{ijkl}_\mathrm{int}\dgD_{(k} \cdot \grd{\displ}_{l)}^\mathrm{int}
    - n_i^\mathrm{ext} \constrel^{ijkl}_\mathrm{ext}\dgD_{(k} \cdot \grd{\displ}_{l)}^\mathrm{ext}\Bigr] \nonumber \\
    - \sigma \Bigl[
      &n_i^\mathrm{int} \constrel^{ijkl}_* n_{(k}^\mathrm{int} \grd{\displ}_{l)}^\mathrm{int}
      - n_i^\mathrm{ext} \constrel^{ijkl}_* n_{(k}^\mathrm{ext} \grd{\displ}_{l)}^\mathrm{ext}\Bigr] \label{eq:numfluxprimal}
  \text{,}
\end{align}
\end{subequations}
where $n_i^\mathrm{ext} = -n_i^\mathrm{int}$ for the purpose of this article.
\Cref{eq:numflux} is the generalized internal-penalty numerical flux defined in
\ccite{dgscheme}, with a choice between $\constrel^{ijkl}_\mathrm{int}$,
$\constrel^{ijkl}_\mathrm{ext}$, and $\constrel^{ijkl}_*$ for every occurrence of the
constitutive relation. The particular choice in \cref{eq:numflux} ensures that
the numerical flux remains consistent, meaning that $(n_i \constrel^{ijkl}
\grd{\strain}_{kl})^* = -n_i^\mathrm{int} \stress^{ij}$ when both $n_i^\mathrm{int}
\constrel^{ijkl}_\mathrm{int}\dgD_{(k} \cdot \grd{\displ}_{l)}^\mathrm{int} = -n_i^\mathrm{ext}
\constrel^{ijkl}_\mathrm{ext}\dgD_{(k} \cdot \grd{\displ}_{l)}^\mathrm{ext} \eqqcolon
-n_i^\mathrm{int} \stress^{ij}$ and $n_{(k}^\mathrm{int}
\grd{\displ}_{l)}^\mathrm{int} = -n_{(k}^\mathrm{ext} \grd{\displ}_{l)}^\mathrm{ext}$. In
particular, note that the penalty term in \cref{eq:numfluxprimal} vanishes when
the displacement is continuous across the boundary, and that the numerical flux
admits solutions where the stress is continuous across the boundary but the
strain is not. Such solutions may arise in a layered material under stress,
because the layers remain \enquote{glued together} but each layer responds to the stress
differently.

The penalty function in \cref{eq:numfluxprimal} is
\begin{equation}\label{eq:penalty}
  \sigma = \penaltyparam \, \frac{(\max(p^\mathrm{int},p^\mathrm{ext})+1)^2}{\min(h^\mathrm{int}, h^\mathrm{ext})}
  \text{,}
\end{equation}
where we make use of the polynomial degree $p$ and a measure of the element
size, $h$, orthogonal to the element boundary on either side of the interface,
as detailed in \ccite{dgscheme}. We choose $C=100$ in this article.

\subsubsection{Boundary conditions}

We impose boundary conditions through fluxes, i.e.\ by a choice of exterior
quantities in the numerical flux~\eqref{eq:numflux}. Specifically, on external
boundaries we set
\begin{subequations}
\begin{align}\label{eq:bc}
  (n_{(k}\grd{\displ}_{l)})^\mathrm{ext} &= (n_{(k}\grd{\displ}_{l)})^\mathrm{int} - 2 n^\mathrm{int}_{(k}\grd{\displ}_{l)}^\mathrm{b} \quad \text{and} \\
  (n_i \constrel^{ijkl} \grd{\strain}_{kl})^\mathrm{ext} &= (n_i \constrel^{ijkl} \grd{\strain}_{kl})^\mathrm{int} + 2 n_i^\mathrm{int} \grd{\stress}^{ij}_\mathrm{b}
  \text{,}
\end{align}
\end{subequations}
where we choose either $\displ^i_\mathrm{b}$ to impose Dirichlet boundary
conditions, or $n_i^\mathrm{int} \stress^{ij}_\mathrm{b}$ to impose Neumann boundary conditions
on the boundary collocation points,
and set the respective other quantity to its interior value.

For the thermal noise problem we impose the pressure induced by the laser beam,
\begin{equation}
  n_i^\mathrm{int} \stress^{ij}_\mathrm{b} = n^j p(r)
  \text{,}
\end{equation}
as Neumann boundary condition on the $z=0$ side of the cylindrical mirror, where
$p(r)$ is the laser beam profile given in \cref{eq:laserbeam}.
On the side of the mirror facing away from the laser we impose
\begin{equation}\label{eq:BCfixed}
  \displ^i_\mathrm{b}=0 \quad \text{(\enquote{fixed})}
\end{equation}
as Dirichlet boundary condition, and on the mantle we impose
\begin{equation}\label{eq:BCfree}
  n_i^\mathrm{int} \stress^{ij}_\mathrm{b}=0 \quad \text{(\enquote{free})}
\end{equation}
as Neumann boundary condition.
\Cref{eq:BCfixed} means that the back of the mirror is held in place, whereas \cref{eq:BCfree}
implies no pressure on the sides, which however, are free to deform in response
to the pressure applied to the front.

\subsection{SpECTRE elliptic solver}

Once discretized, the linear algebraic equations~\eqref{eq:dg_residuals} are
solved numerically for the displacement vector field values~$\grd{\displ}^i$ on all
elements and grid points in the computational domain. As is typical for
discretized elliptic equations, \cref{eq:dg_residuals} defines a matrix equation
\begin{equation}\label{eq:matrix_eqn}
  \opA \grd{\displ} = \grd{b}
  \text{,}
\end{equation}
where $\grd{\displ}$ denotes the set of all $N_\mathrm{DOF}=3 \times
N_\mathrm{points} = 3 \times \sum_k N_k$ displacement vector field values in the
computational domain, and $\opA$ is a matrix with $N_\mathrm{DOF} \times
N_\mathrm{DOF}$ entries. To solve \cref{eq:matrix_eqn} means inverting the
matrix $\opA$. However, as the resolution of the computational domain increases,
the matrix $\opA$ easily becomes too large to construct explicitly, to store on
an ordinary computer, and to invert directly.

Therefore, we solve \cref{eq:matrix_eqn} with the elliptic solver component of
the open-source \spectre{} code~\cite{spectre,ellsolver}. It employs an
iterative generalized minimal residual (GMRES) algorithm to solve
\cref{eq:matrix_eqn} to the requested precision. A multigrid preconditioner
accelerates the GMRES algorithm by supporting each iteration with an approximate
solution from a hierarchy of successively coarser grids. On every grid, an
additive Schwarz smoother decomposes the problem into many overlapping
subproblems, one per element in the domain, which are solved independently and
in parallel. The subproblems are distributed across the cores of a computing
cluster by a task-based parallelization paradigm. The elliptic solver is
described in detail in \ccite{ellsolver}.

\section{Results}\label{sec:results}

\begin{figure*}
  \centering
  \includegraphics[width=2.\figwidth]{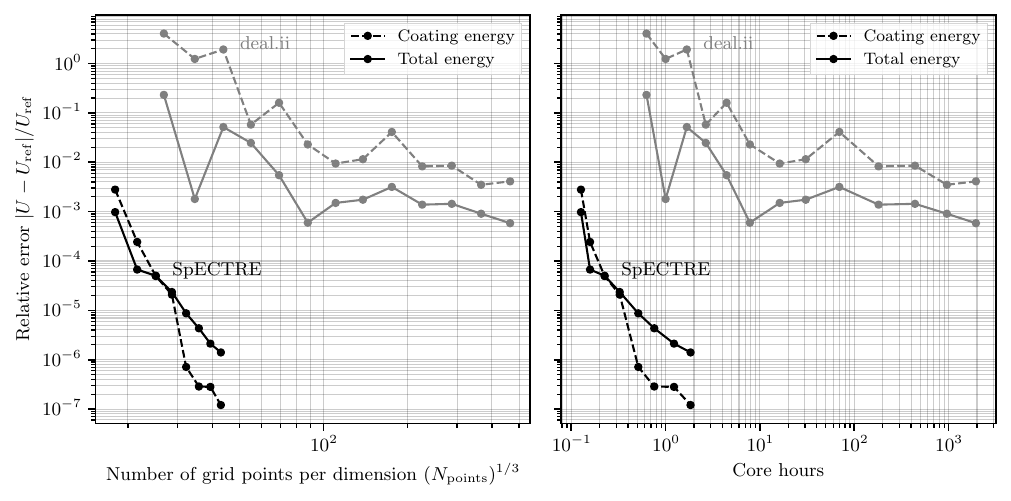}
  \caption{
    \label{fig:comparison}
    Relative error of the potential energy in a single amorphous coating layer
    (dashed lines) and in the full mirror (solid lines). \emph{Left:}
    \spectre{}, with our discontinuous Galerkin method, resolves the coating
    layer at high precision using only a fraction of the number of grid points
    needed by the \dealii{} approach. \emph{Right:} \spectre{} solves the
    elliptic problem using only a fraction of the computational resources needed
    by the \dealii{} approach.}
\end{figure*}

Our simulations with \spectre{} were performed on one or more 16-core compute nodes,
each with \SI{64}{\giga\byte} of memory and two eight-core Intel Haswell
E5-2630v3 processors clocked at \SI{2.40}{\giga\hertz}, connected with an Intel
Omni-Path network. We distribute the elements that compose the computational
domain evenly among cores, leaving one core per node free to perform
communications.

We compare our results to previous work using an open-source
finite element code to calculate the Brownian coating thermal noise for
amorphous and crystalline materials. Its methods are described in
Secs. 2.4--2.6 of \ccite{Lovelace2017xyf}. The code was built using the
\dealii{}~\cite{dealii, dealii2019design} finite element framework and we
henceforth refer to it as \dealii{}. It adopted a standard weak form of the
elastostatic equations, discretized them using a conventional finite element
approach, and solved them using \dealii{} with the \texttt{PETSc}~\cite{petsc}
conjugate gradient linear solver and the \texttt{ParaSAILS} preconditioner in
the \texttt{hypre}~\cite{hypre} package. The \dealii{} code relies on the
Message Passing Interface (MPI) for parallelization.

\subsection{Single-coating comparison}\label{sec:comparison}

First, we consider the single-coating scenario investigated in
\ccite{Lovelace2017xyf} and demonstrate the superior performance of our new
approach. We choose the parameters listed in \ccite{Lovelace2017xyf}, Table 1
for a cylindrical mirror of radius $R=\SI{12.5}{\milli\metre}$ with a single
$d=\SI{6.83}{\micro\metre}$ thin effective-isotropic \algaas{} coating. We
simulate the scenario both with the \dealii{} approach employed in
\ccite{Lovelace2017xyf} and with our new approach with
the \spectre{} code.

\Cref{fig:comparison} presents the numerical precision and computational cost of
both approaches. To assess the numerical precision we successively increase the
resolution in both codes. In \spectre{} we increase the resolution by
incrementing the number of grid points in all dimensions of all elements in the
domain by one, and in \dealii{} we employ an adaptive mesh-refinement scheme~\cite{Lovelace2017xyf}.
We compute the error in the elastic potential energy
relative to a high-resolution reference configuration simulated in \spectre{},
for which we have split all elements of the highest-resolution configuration
included in \cref{fig:comparison} in half along all three dimensions (see also
\cref{fig:domain}). Hence, the high-resolution reference configuration has $\sim
79$ grid points per dimension.

We find that both codes converge to the same
solution, but our new approach in \spectre{} achieves about four orders of
magnitude higher accuracy than the \dealii{} approach using the same number of
grid points. Furthermore, our new approach simulates this scenario with
sub-percent error in only \SI{30}{\second} on \num{15} cores, for which the
\dealii{} approach required multiple hours on \num{324} cores.
Our new approach also achieves a fractional error below
$10^{-5}$ in only half a core-hour, or two minutes of real time, which was
prohibitively expensive with the \dealii{} approach.

\subsection{Accuracy of the approximate analytic solution}

\begin{figure}
  \centering
  \includegraphics[width=\figwidth]{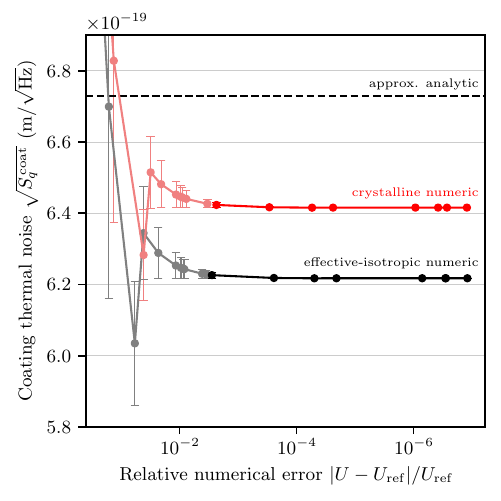}
  \caption{
    \label{fig:thermalnoise}
    Thermal noise in an \algaas{}-coated mirror computed from the approximate
    analytic solution~\eqref{eq:approx_analytic} and from our numerical
    simulations. The effective-isotropic simulation (black) retains the
    amorphous approximation for the material, but includes finite-size effects.
    The crystalline simulation (red) eliminates this approximation. Previous
    simulations with the \dealii{} approach are shown in lighter colors to the
    left.}
\end{figure}

Second, we study the accuracy of the approximate analytic solution for the
single-coating thermal noise, \cref{eq:approx_analytic}, using the superior
numerical precision we can now achieve over the results presented in
\ccite{Lovelace2017xyf}. The approximate solution holds for a thin coating, $d
/ r_0 \ll 1$, a semi-infinite mirror, $r_0 / R \ll 1$ and $d / R \ll 1$, and for
isotropic-homogeneous materials. Therefore, it does not capture the finite-size
effects included in our simulations, and approximates the crystalline \algaas{}
coating as an amorphous material.

To assess the magnitude of the finite-size effects, we employ the simulations
detailed in \cref{sec:comparison}, which use the same effective-isotropic model
for the \algaas{} coating that underpins the approximate analytic solution.
\Cref{fig:thermalnoise} presents both the thermal noise computed from the
simulations and the approximate analytic solution (black). Error bars are computed as
$\Delta \sqrt{S_q^{\,\coat}} / \sqrt{S_q^{\,\coat}} = 1/2 \, \Delta U_\coat / U_\coat$ from the relative numerical
error in the elastic potential energy. While \ccite{Lovelace2017xyf} estimated
the magnitude of finite-size effects for this problem to \SI{7}{\percent}, we
can now report that their simulations captured the effect to \SI{7.5 \pm
0.2}{\percent}. With our new numerical method, we can make this statement more
precise and report a finite-size effect of \SI{7.616649 \pm 0.000006}{\percent}.

To assess the magnitude of the amorphous approximation to the crystalline
coating material, we repeat the simulations with a crystalline constitutive
relation and the parameters listed in \ccite{Lovelace2017xyf}, Table 1.
The thermal noise computed from these simulations is presented in
\cref{fig:thermalnoise} as well (red). We refine the estimate of \SI{4}{\percent}
from \ccite{Lovelace2017xyf} to \SI{4.5 \pm 0.2}{\percent}, and report
\SI{4.667990 \pm 0.000006}{\percent} using our new numerical method.

Note that we report only numerical errors from our simulations here. However,
the parameters that define our simulations, such as coating and substrate
material properties, are typically measured experimentally and carry significant
uncertainties. For example, elastic moduli recently reported in
\ccite{Amato:2021ryz} were measured on the percent level. Computing the thermal
noise by \cref{eq:thermalnoise} also involves loss angles~$\lossang$ which can
be measured on the percent level as well~\cite{Amato:2021ryz}. Therefore,
computational resources are spent most effectively to drive numerical errors
below sub-percent levels and no further. With our new numerical methods we
achieve sub-percent accuracy with a fraction of the computational resources
required before (see \cref{sec:comparison}).

\subsection{Multiple sub-wavelength crystalline coatings}

\begin{figure*}
  \centering
  \includegraphics[width=2.\figwidth]{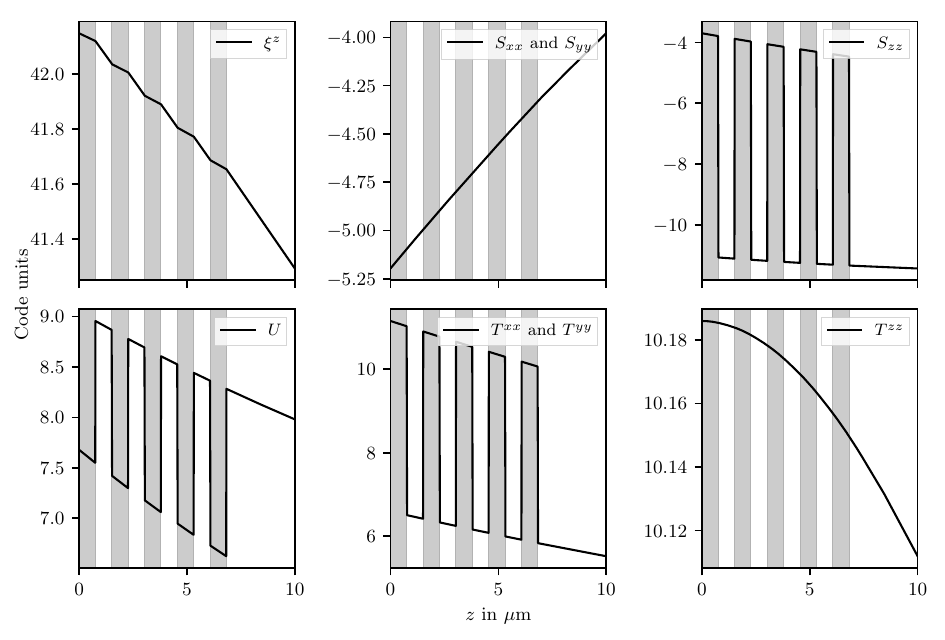}
  \caption{
    \label{fig:multicoating_vars}
    Elastic variables along the $z$-axis for a mirror with multiple thin coating
    layers, simulated with \spectre{}. Gray layers are crystalline \algaas{},
    and white regions are fused silica. The nine coating layers have a combined
    thickness of \SI{6.83}{\micro\metre} and the material extends to
    $z=\SI{12.5}{\milli\metre}$ outside the range of this plot.}
\end{figure*}

Finally, we apply our new computational approach to a scenario that presents many
of the challenges we expect for applications to realistic mirror configurations.
We simulate a cylindrical mirror of the same radius $R=\SI{12.5}{\milli\metre}$
as before, but split the $d=\SI{6.83}{\micro\metre}$ thin coating into nine
layers, so the thickness of each coating layer is below the typical
\SI{1}{\micro\metre} wavelength of the laser. The coating layers alternate
between fused silica and crystalline \algaas{}, with the elastic moduli
$c_{11}$, $c_{12}$ and $c_{44}$ listed in \ccite{Lovelace2017xyf}, Table 1.
Neither sub-wavelength coatings nor multiple layers were simulated in
\ccite{Lovelace2017xyf}, but our new computational approach in \spectre{}
achieves both.

\begin{figure}
  \centering
  \includegraphics[width=\figwidth]{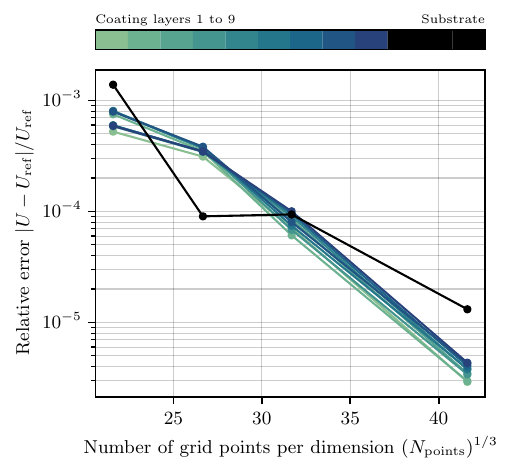}
  \caption{
    \label{fig:multicoating_convergence}
    Convergence test of the elastic potential energy in each coating layer,
    and in the substrate. Our new numerical method achieves exponential
    convergence despite the discontinuous material properties.}
\end{figure}

\Cref{fig:multicoating_vars} presents our numerical solution of this scenario.
Our new computational approach based on discontinuous Galerkin methods resolves
the thin coating layers at high accuracy without spurious oscillations.
\Cref{fig:multicoating_convergence} presents the numerical precision of the
solution. We increase the resolution by incrementing the number of grid points
per element and dimension and compute the relative error to a high-resolution
reference configuration, as we did in \cref{sec:comparison}. The error converges
exponentially, which is a feature of our discontinuous Galerkin method with grid
boundaries placed at the layer interfaces.

\section{Discussion}\label{sec:conclusion}

We have presented a new numerical method to model Brownian thermal noise in thin
mirror coatings based on a discontinuous Galerkin (DG) discretization. With our
new method, we model thermal noise in a one-inch cylindrical mirror with a
microns-thick coating at unprecedented accuracy at a fraction of the time needed
in a previous, conventional finite-element approach~\cite{Lovelace2017xyf}.
Using these high-accuracy simulations, we find that a commonly-used approximate
analytic solution overestimates the coating thermal noise for this problem by
\SI{7.6}{\percent} when taking only finite-size effects into account, and by
\SI{4.7}{\percent} when modeling it as a crystalline material, which refines a
previous estimate in \ccite{Lovelace2017xyf}. We also demonstrate that, unlike
the approach in Ref.~\cite{Lovelace2017xyf}, our new method is capable of
resolving multiple sub-wavelength coatings, including coatings of a
cubic-crystalline material. Our new numerical method is implemented in the
open-source \spectre{} code and the results presented in this article are
reproducible with the supplemental input-file configurations.

We found that it is crucial for the success of our new method that the interfaces
between layers of different materials coincide with element boundaries in our
computational domain. Then, our discontinuous Galerkin discretization with a
suitable choice of numerical flux converges exponentially, achieving high
accuracy with a small number of grid points. The scheme can potentially be
improved in future work. Most notably, an adaptive mesh-refinement (AMR)
algorithm would have great potential to further improve the accuracy and
efficiency of the scheme, by distributing the resolution in the computational
domain to regions and dimensions where it is most needed.

Furthermore, the elliptic solver in the \spectre{} code that we employ to solve
the discretized problem numerically can be improved to accelerate thermal-noise
calculations. The calculations we have presented in this article require a few
hundred solver iterations to converge, or up to $\sim 1400$ for our
highest-resolution simulation with multiple sub-wavelength crystalline coatings.
While simple configurations complete in seconds or minutes of real-time on
\num{15} cores, where the previous approach needed hours on \num{324} cores, the
more challenging configurations, which were prohibitively expensive with the
previous approach, solve in about an hour on \num{45} cores.

We expect additional speedup with further improvements to the
elliptic solver algorithm in \spectre{}. In particular, improvements to its
multigrid preconditioner have great potential to speed up the simulations. The
multigrid algorithm relies on solving the problem
approximately on coarser grids to resolve large-scale modes in the solution. It
currently cannot coarsen the grid any further than the size of each coating
layer because the layers define the material properties. To accelerate the
calculations, we intend to let the multigrid algorithm combine layers with
different materials into fiducial coarse layers with effective material
properties. This approach is possible because the partitioning of the domain
into layers is necessary only to define material properties, not to define the
cylindrical shape of the domain. \Ccite{ellsolver} shows that the multigrid algorithm can
achieve resolution-independent iteration counts when the domain can be coarsened
sufficiently. Note that the fiducial coarse layers affect only the convergence
speed of the solver and do not change the solution once the solver has
converged.

Our numerical models of thermal noise have the potential to inform upgrades that
increase the sensitivity of gravitational-wave detectors, using the advanced
computational technology that we develop for numerical-relativity simulations in
the \spectre{} code.
In the future, we intend to apply our new numerical method to simulate Brownian
thermal noise in more realistic mirror configurations and materials that are
under consideration for current and future gravitational-wave detectors, such as
the optimized configuration found in \ccite{venugopalan2021global}. While
approximate analytic solutions can provide useful estimates, only numerical
models can precisely quantify the finite-size effects of changing the mirror
geometry. In particular, finite-size effects are more important for real
gravitational-wave detectors than for tabletop experiments measuring thermal noise.
Tabletop experiments often use small beam sizes to enlarge the thermal noise and
hence make it easier to measure, whereas gravitational-wave detectors prefer
large beam sizes to minimize thermal noise. Therefore, we plan to employ
our new numerical method to explore realistic mirror configurations, with the
goal of finding configurations that minimize Brownian coating thermal noise.

\ack
The authors thank Josh Smith for helpful discussions. Computations were
performed with the \spectre{}~\cite{spectre} and \dealii{}~\cite{dealii, dealii2019design} codes
on the \texttt{Minerva} cluster at the Max Planck Institute for Gravitational
Physics and on the \texttt{Ocean} cluster at Fullerton. The figures in this
article were produced with \texttt{dgpy}~\cite{dgpy}, \texttt{matplotlib}~\cite{matplotlib1,matplotlib2},
\texttt{TikZ}~\cite{tikz} and \texttt{ParaView}~\cite{paraview}.
This work was supported in part by the Sherman Fairchild Foundation, by NSF
Grants No.\ PHY-1654359 and No.\ AST-1559694 at Cal State Fullerton, by NSF
Grants No.\ PHY-2011961, No.\ PHY-2011968, and No.\ OAC-1931266 at Caltech and
by NSF Grants No.\ PHY-1912081 and No.\ OAC-1931280 at Cornell.

\section*{References}

\bibliographystyle{iopart-num}
\bibliography{References}

\end{document}